\documentclass[lettersize,journal]{IEEEtran}
\usepackage[utf8]{inputenc}
\usepackage[T1]{fontenc}
\usepackage{amsmath,amssymb,amsthm}
\usepackage{amsfonts,bm}
\usepackage{graphicx}
\usepackage[export]{adjustbox}
\usepackage[font=footnotesize]{caption}
\usepackage[caption=false,font=normalsize,labelfont=sf,textfont=sf]{subfig}
\usepackage{textcomp,stfloats,url,verbatim,booktabs}
\usepackage[boxed,ruled,vlined,linesnumbered]{algorithm2e}
\usepackage{array,tabularx,boldline,diagbox,colortbl}
\usepackage{multirow,makecell,longtable}
\usepackage{lettrine,lipsum}
\usepackage{hyperref}
\hypersetup{colorlinks=true,linkcolor=blue,urlcolor=blue,citecolor=blue}

\newtheorem{theorem}{Theorem}
\newtheorem{lemma}{Lemma}
\newtheorem{corollary}{Corollary}
\newcommand{\reffig}[1]{Fig.~\ref{#1}}
\newcommand{\reftable}[1]{Table~\ref{#1}}
\newcommand{\refeq}[1]{Eq.~\ref{#1}}

\def\BibTeX{{\rm B\kern-.05em{\sc i\kern-.025em b}\kern-.08em
    T\kern-.1667em\lower.7ex\hbox{E}\kern-.125emX}}
\usepackage{balance}

\begin{document}

\title{TextCrafter: Optimization-Calibrated Noise for Defending Against Text Embedding Inversion}
\author{Duoxun Tang, Xinhang Jiang and Jiajun Niu
  \thanks{Duoxun Tang is with Shenzhen International Graduate School, Tsinghua University, China. Email: tdx25@mails.tsinghua.edu.cn.
  Xinhang Jiang is with the School of Data Science, The Chinese University of Hong Kong, Shenzhen, China. Email: xinhangjiang@link.cuhk.edu.cn.
  Jiajun Niu is with Shenzhen International Graduate School, Tsinghua University, China. Email: njj25@mails.tsinghua.edu.cn.
  }
}

\markboth{Journal of \LaTeX\ Class Files,~Vol.~18, No.~9, September~2020}%
{How to Use the IEEEtran \LaTeX \ Templates}

\maketitle

\begin{abstract}
Text embedding inversion attacks reconstruct original sentences from latent representations, posing severe privacy threats in collaborative inference and edge computing. We propose \textbf{TextCrafter}, an optimization-based adversarial perturbation mechanism that combines RL-learned, geometry-aware noise injection orthogonal to user embeddings $\mathbf{f}_0$ with cluster priors $\mathbf{f}_u$ and PII-signal guidance to suppress inversion while preserving task utility. Unlike prior defenses—either non-learnable or agnostic to perturbation direction—TextCrafter provides a directional protective policy that balances privacy and utility. Under strong privacy (\textsc{bleu}$<$3, \textsc{rouge-l}$<$15), TextCrafter maintains $\geq$70\% classification accuracy on four datasets and consistently outperforms Gaussian/LDP baselines across lower privacy budgets, demonstrating a superior privacy-utility trade-off.

\end{abstract}


\section{INTRODUCTION}

\begin{figure}
\centering
	\captionsetup{
		font={scriptsize}, 
	}
	\begin{adjustbox}{valign=t}
		\includegraphics[width=0.95\linewidth]{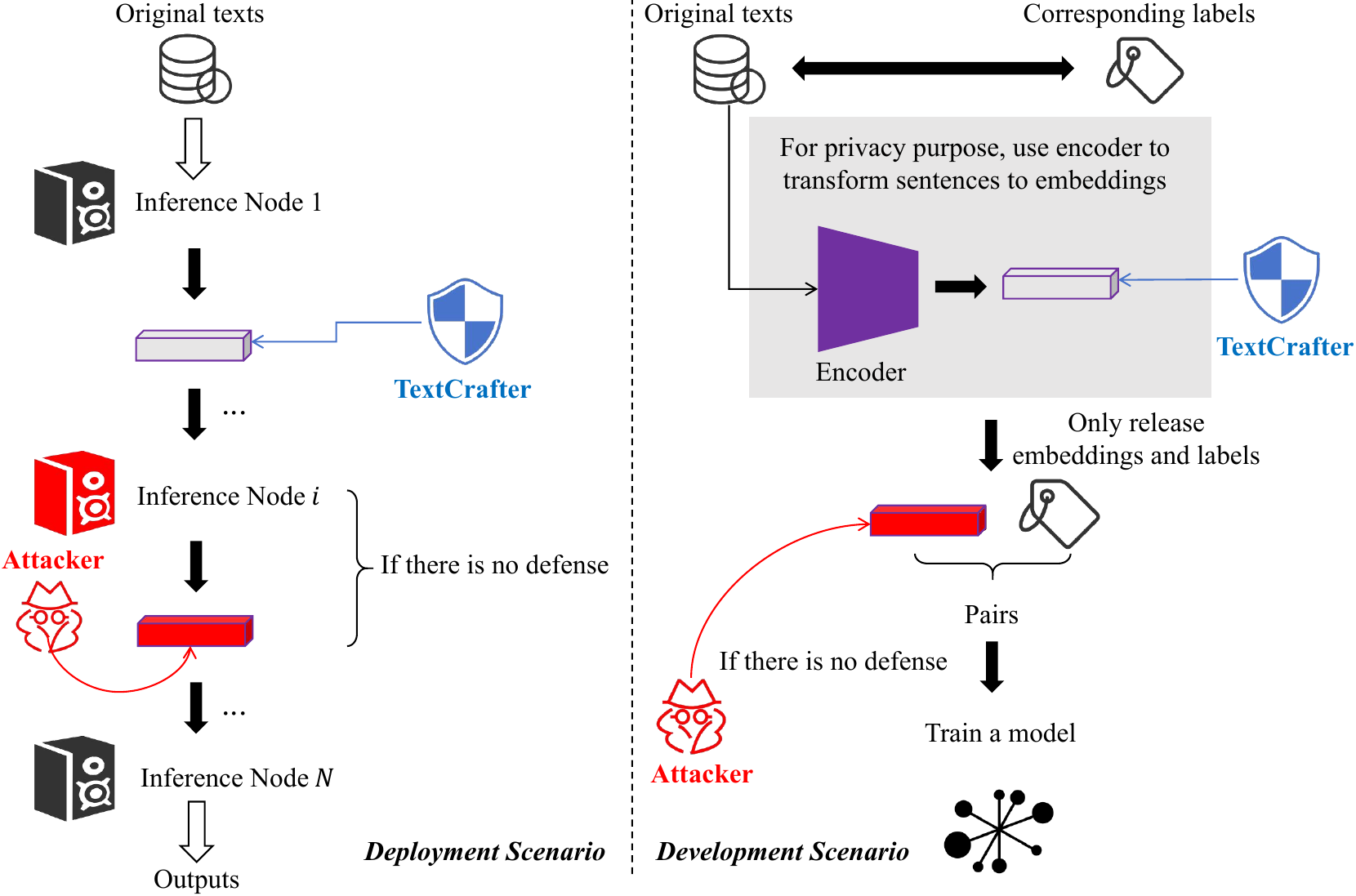}
	\end{adjustbox}
	\caption{Examples of when TextCrafter is needed in two typical scenarios. On the left, protected embeddings are fed to a cloud-based, pre-trained model for inference without ever exposing raw text; on the right, a fresh classifier is trained solely on vectors and labels, still oblivious to the original sentences. One policy stops snooping in both cases, keeping privacy and utility.}
	\label{figs:abstract}
    \vspace{-4mm}
\end{figure}

\lettrine{W}{ith} the proliferation of large language models (LLMs), flagship systems now permeate domains ranging from medical triage \cite{che2025llm} to on-device personal assistants \cite{appleSiri2025}.  However, their hundreds of billions of parameters make standalone deployment prohibitive on edge System-on-Chips (SoCs).  Split-compute pipelines—where a smartphone encodes inputs into embeddings and uploads only the vector to a cloud accelerator—have become the de-facto solution for real-time inference, but they expose intermediate representations to inversion attacks that can reconstruct original sentences containing personal-identifiable information (PII) or confidential enterprise data (\reffig{figs:abstract}, deployment scenario). Beyond inference, protected embeddings are used from the very beginning of the training pipeline: the cloud provider releases only perturbed representations, and the downstream partner train its model on these embeddings (\reffig{figs:abstract}, development scenario).

Embedding-inversion attacks were early explored in computer vision \cite{mahendran2015understanding}, where deep features could be reversed to reveal the original image.  Recent studies show that sentence embeddings are equally vulnerable \cite{morris2023text}: given only black-box access to an LLM encoder and a few thousand queries, an adversary can train a surrogate inverter that accurately reconstructs the original text.  Once the sentence is recovered, harms are immediate—patient names in clinical notes, deal amounts in financial wires, or personal dialogues from on-device assistants.  

To counter inversion attacks while preserving downstream utility, prior studies adopt perturbation-based defenses: Morris et al. \cite{morris2023text} inject Gaussian noise into embeddings, and Chen et al. \cite{chen2024text} extend this approach to multilingual settings, observing reduced effectiveness beyond English. Differential Privacy (DP) \cite{dwork2014algorithmic} bounds the influence of any single sample and has been shown to conceal gradient-level signals when applied during training; sequence-level metric local DP (metric-LDP) \cite{du2023sanitizing} can also be enforced at inference by perturbing the entire sentence vector. Existing defenses either inject one-shot Gaussian noise—soon filtered by adaptive attackers—or enforce LDP with isotropic perturbations, which still degrades utility and ignores the embedding geometry. We bridge this gap with \textbf{TextCrafter}, a lightweight, optimization-based protector that learns to plant noise orthogonal to sensitive directions, preserving utility while blocking inversion in a single forward pass.

To close the above gaps we introduce \textbf{TextCrafter}, a lightweight, optimization-based defence that converts the embedding into a privacy--utility controllable interface. Concretely, TextCrafter upgrades Gaussian/LDP's one-shot blind noise into a learnable, geometry-aware protective perturbation: leveraging attention-based direction learning and RL—guided by cluster priors and trained PII classifier that supplies PII signals. It optimises a directed perturbation that minimises PII leakage and semantic drift in a single forward pass, something unachievable with Gaussian/LDP's fixed, non-learnable noise. Extensive experiments across four datasets against Vec2Text attack show TextCrafter achieves stronger privacy and higher utility than existing perturbation defenses, validating its practicality for real-world application.

\begin{figure*}[t]
\begin{center}
  \includegraphics[width=0.8\linewidth]{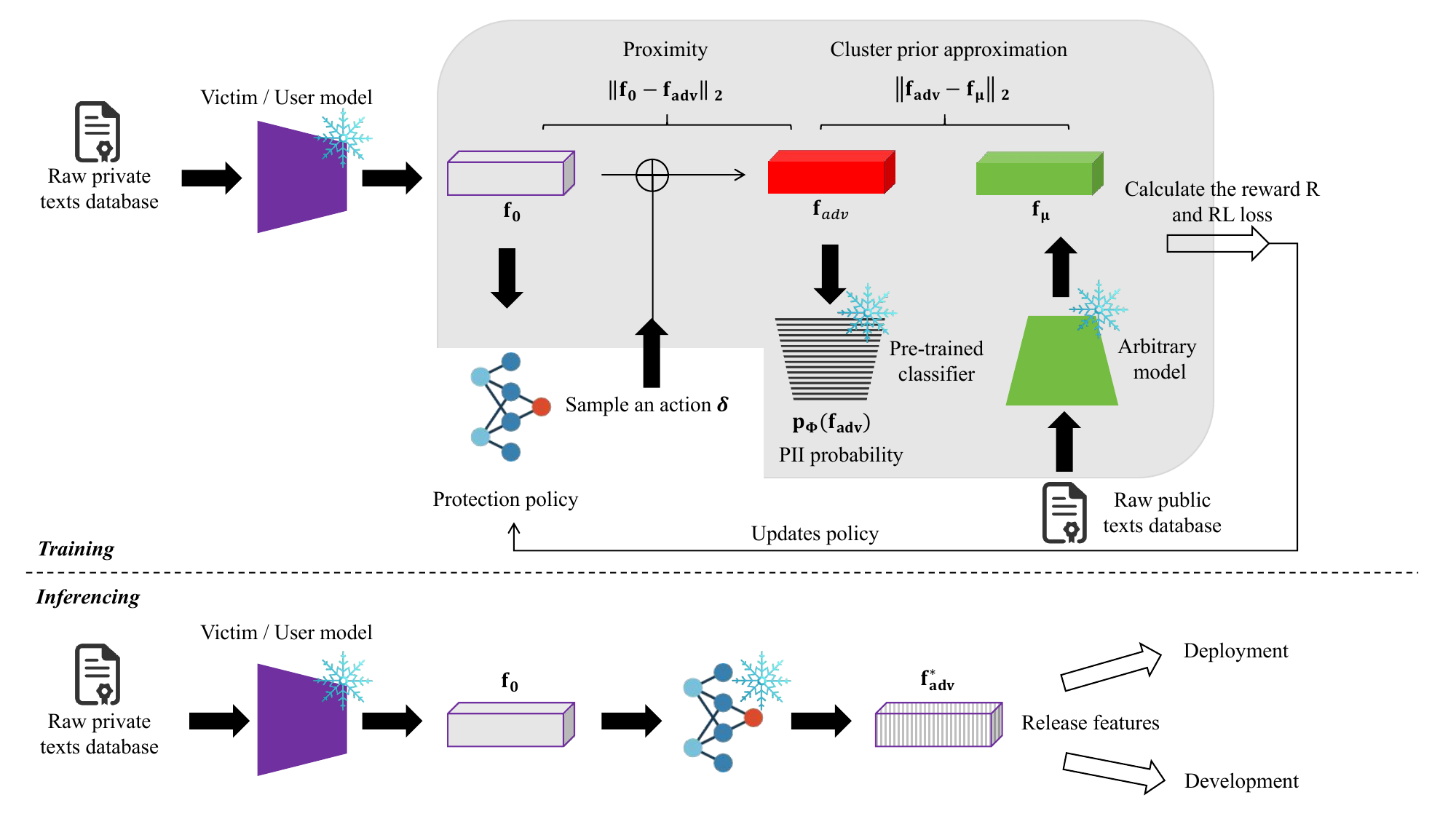}
\end{center}
  \caption{Architecture of \textbf{TextCrafter}: TextCrafter employs a learnable policy to generate reasonable, protective perturbations; the snowflake symbol indicates frozen weights. Explanations of the remaining symbols are provided in the Method section.}
\label{fig:framework}
\vspace{-3mm}
\end{figure*}

\section{PRELIMINARY}
\subsection{Security Setting}
\subsubsection{Attacker’s Knowledge}
The adversary has access to the publicly released embedding $\mathbf{f}_{0}$; hence attacker can compute $\mathbf{f}_{0}$ from any observed text and, conversely, run unlimited inversion queries to recover a high-fidelity text reconstruction $\mathbf{x}^{*}$ from $\mathbf{f}_{0}$ such as black-box attack from Vec2Text \cite{morris2023text}.

\subsubsection{Defender’s Knowledge}
The defender controls the clean embedding $\mathbf{f}_{0}$ before release.  Defender can collect public datasets to train a privacy-identifiable-information (PII) classifier, denoted by $p_{\Phi}(\cdot)$, and construct a private (local) network to learn how to add protective perturbations, both of which remain unavailable to the attacker.

\subsection{Defense Scenarios}
In the deployment scenario, protected embeddings are fed into an already-trained cloud LLM for inference; A realistic case is collaborative LLM inference \cite{jiang2023hexgen}: the edge device encodes the prompt, and an attacker with physical or Direct Memory Access (DMA) access can steal the tensor and run inversion attack.  In the development scenario, protected embeddings are used from the very beginning of the training pipeline: the cloud provider releases only perturbed representations, and the downstream partner train its model on these embeddings. Here the trade-off is straightforward: stronger privacy perturbation lowers the risk of sentence reconstruction, but also risks degrading the model’s downstream accuracy.

\subsection{Problem Definition}
An \emph{inversion attack} takes a publicly released embedding
$\mathbf{f_0}\in\mathbb{R}^{D}$ and feeds it to an adversary's decoder
$\mathcal{A}:\mathbb{R}^{D}\to\mathcal{T}$ (where $\mathcal{T}$ is the text
space) to recover the original sentence $x$ or a semantically equivalent
variant $x'$.  Formally, the adversary succeeds when
\begin{equation}
\mathsf{Sim}(x,\mathcal{A}(\mathbf{f_0}))\ge\tau,\qquad\tau>0,
\end{equation}
for some similarity metric $\mathsf{Sim}$ (e.g., BLEU, token-F1).

To thwart such attacks, we wish to publish a \emph{protected} embedding
$\mathbf{f}_{\mathrm{adv}}$ that satisfies two conflicting goals:

\begin{enumerate}
  \item[(i)] \textbf{Privacy:}  inversion quality must drop, i.e.\
  \begin{equation}
  \mathsf{Sim}(x,\mathcal{A}(\mathbf{f}_{\mathrm{adv}}))\le\tau-\Delta,\qquad\Delta>0.
  \end{equation}

  \item[(ii)] \textbf{Utility:}  downstream task performance must remain intact, i.e.\
  \begin{equation}
  \|\mathbf{f}_{\mathrm{adv}}-\mathbf{f_0}\|_{2}\le\varepsilon\quad\text{and}\quad
  \mathsf{Task}(\mathbf{f}_{\mathrm{adv}})=\mathsf{Task}(\mathbf{f_0}),
  \end{equation}
\end{enumerate}
where $\varepsilon>0$ is a user-chosen bound and $\mathsf{Task}$ denotes any
application-specific evaluation (classification accuracy, retrieval rank,
etc.).

The challenge is to find a \emph{release mapping}
\begin{equation}
\mathcal{M}:\mathbb{R}^{D}\to\mathbb{R}^{D},\qquad
\mathbf{f}_{\mathrm{adv}}=\mathcal{M}(\mathbf{f_0})
\end{equation}
such that for \emph{every} embedding $\mathbf{f_0}$ goals (i)–(ii) hold
\emph{deterministically}.

\subsection{Datasets}
We evaluate different defense methods on Financial PhraseBank, ADE, SST-2 and AG News; all four datasets serve as test-beds for sentiment or relation classification under privacy-preserving embeddings. And corporate WikiText-2 to train PII classifier.

\subsection{Metrics and Baselines}
\textbf{Privacy} is quantified by five metrics. BLEU, Token-F1, Exact-Match (the proportion of reconstructed sentences that are identical to the original), Cosine and Rouge-L. \textbf{Utility} is assessed through two metrics. Macro-F1 and CLS-ACC. We compare TextCrafter with three defenses: \textbf{Shuffling} (random permutation of $\mathbf{f}_0$ along the hidden dimension), \textbf{Gaussian} (isotropic noise $\mathcal{N}(0,\varepsilon^2\mathbf{I})$, $\varepsilon\!\in\!\{0.001, 0.005, 0.01, 0.05, 0.1, 0.5, 1\}$), and \textbf{LDP} (on-device Gaussian mechanism yielding $(\varepsilon,10^{-5})$-LDP on unit-sphere embeddings, $\varepsilon\!\in\!\{1,4,8,12\}$).

\section{METHOD}

\subsection{Design of TextCrafter}

\subsubsection{PII-Classifier} 

We first define two corpora $\mathcal{D}_{\mathrm{pii}}$ and $\mathcal{D}_{\mathrm{plain}}$.
From each corpus we draw $n$ sentences:

\begin{equation}
\mathcal{S}_{\mathrm{pii}}\sim\mathcal{D}_{\mathrm{pii}},\quad
\mathcal{S}_{\mathrm{plain}}\sim\mathcal{D}_{\mathrm{plain}},\quad
|\mathcal{S}_{\mathrm{pii}}|=|\mathcal{S}_{\mathrm{plain}}|=n.
\label{5}
\end{equation}

An open-source data-protection toolkit \cite{presidio} $\mathrm{PII}(\cdot)$ re-labels the plain split:
\begin{align}
\mathcal{S}_{\mathrm{pii}}' &=
\mathcal{S}_{\mathrm{pii}}\cup
\{\bm{x}\in\mathcal{S}_{\mathrm{plain}}\mid\mathrm{PII}(\bm{x})=1\},\\[4pt]
\mathcal{S}_{\mathrm{plain}}' &=
\mathcal{S}_{\mathrm{plain}}\setminus
\{\bm{x}\in\mathcal{S}_{\mathrm{plain}}\mid\mathrm{PII}(\bm{x})=1\}.
\label{eq:set_definition}
\end{align}

Intuitively, above equations moves any sentence mistakenly labeled as non-PII back into the PII set and removes all confirmed PII sentences from the original non-PII set. Together, these two steps align the training labels with the true distribution and naturally yield an approximately 1:1 balanced sample. The resulting label ratio eliminates class imbalance:

\begin{equation}
\rho=\frac{|\mathcal{S}_{\mathrm{pii}}'|}{|\mathcal{S}_{\mathrm{pii}}'|+|\mathcal{S}_{\mathrm{plain}}'|}\approx 0.5,
\end{equation}

A logistic-regression classifier $p_{\Phi}(\cdot)$ is trained on the pooled embeddings $\bm{e}$ produced by a frozen encoder. And standard metrics are evaluated on a held-out set $\mathcal{S}_{\mathrm{val}}$.

\begin{equation}
p_{\Phi}\colon\mathbb{R}^{d}\to[0,1],\qquad
\hat{y}=p_{\Phi}(\bm{e}).
\end{equation}

\subsubsection{Protect Policy Network}
The policy network, dubbed ProtectPolicy, consists of two symmetric branches built upon an identical transformer block. Each block is a single-layer multi-head attention (MHA) \cite{vaswani2017attention} module followed by a feed-forward network (FFN). Formally, for an input embedding $x \in \mathbb{R}^{1 \times D}$ we first add a learnable positional vector $p \in \mathbb{R}^{1 \times D}$ and obtain
$h = x + p$. The block then applies multi-head self-attention, which can be formulated as follows:
\begin{equation}
\mathbf{A}=\mathrm{MHA}(\mathbf{h},\mathbf{h},\mathbf{h})
=\mathrm{Concat}\bigl(\mathrm{head}_{1},\dots,\mathrm{head}_{H}\bigr)\mathbf{W}^{O}
\end{equation}

\begin{equation}
\mathrm{head}_{h}= \mathrm{Softmax}\!\left(
\frac{\mathbf{h}\mathbf{W}^{Q}_{h}(\mathbf{h}\mathbf{W}^{K}_{h})^{\!\top}}{\sqrt{d_{k}}}
\right)\!\mathbf{h}\mathbf{W}^{V}_{h},
\end{equation}
where
$\mathbf{W}^{Q}_{h},\mathbf{W}^{K}_{h}\in\mathbb{R}^{D\times d_{k}}$,
$\mathbf{W}^{V}_{h}\in\mathbb{R}^{D\times d_{v}}$,
$\mathbf{W}^{O}\in\mathbb{R}^{H d_{v}\times D}$,
and
$H=12$,
$d_{k}=d_{v}=\frac{D}{H}=64$,
$\mathbf{h}\in\mathbb{R}^{B\times 1\times D}$,
with dropout $0.1$.
After that, A residual connection and LayerNorm yield:
\begin{equation}
\mathrm{h'} = \mathrm{LayerNorm(h + Dropout(A))}.
\end{equation}
Subsequently, a two-layer FFN with GELU activation and dropout is employed:
\begin{equation}\mathrm{F = Linear(GELU(Linear(h')))}.  \end{equation}
\begin{equation}\mathrm{y = LayerNorm(h' + Dropout(F))}.\end{equation}
The first branch uses y to predict the mean of the raw noise $\delta$ via a linear layer. The second branch shares the same architecture but independently produces a direction vector v.

\subsubsection{RL-based Optimization}
We design a PII-aware instant embedding defense via RL, named TextCrafter. Specifically, we define an agent entity named ProtectPolicy, whose responsibility is to output an action (perturbation) $a=\delta$ and its log-probability log\_prob given the current state (clean user inputs) $s=\mathbf{f}_{0}$. The policy network $\pi(a|s)$ is a continuous Gaussian policy and is updated via policy gradient. Below, we describe three aspects: construct the action distribution, add the perturbation, and the design of the reward function.

\textbf{Construct the action distribution.}
By learning the mean $\mu$ and standard deviation $\sigma$ of a Gaussian distribution, we construct an action-sampling distribution $\mathcal{N}(\mu, \sigma^{2})$. Concretely, $\mu$ is obtained by adding a learnable positional encoding to the input $\mathbf{f}_0$, feeding the result into a multi-head attention block mentioned above, and passing the output through a final linear layer. The standard deviation $\sigma$ is directly modeled as a learnable parameter. Thus we build a Gaussian policy with mean $\mu$ and variance $\sigma^{2}$, from which the perturbation $\delta$ is sampled. For clearer exposition, the formulation is given as follows.
\begin{equation}
\begin{aligned}
\boldsymbol{\mu} &=\mathbf{W}_{\mu}\,\mathrm{MHA}\bigl(\mathbf{f}_{0}+\mathbf{p}\bigr),&
\boldsymbol{\sigma} &=e^{\boldsymbol{\lambda}},&
\boldsymbol{\delta} &\sim\mathcal{N}(\boldsymbol{\mu},\,\boldsymbol{\sigma}^{2}\mathbf{I}).
\end{aligned}
\end{equation}
Here $\mathbf{W}_{\mu}$ is the linear year, $\mathbf{p}$ denotes the learnable positional encoding and $\lambda$ represents the learned log-standard-deviation parameters.

\textbf{Add the perturbation.}
Similarly, we use a learnable network to produce the reference direction $v$ for protective perturbations; as shown in \refeq{eq:unit}, we obtain the unit direction vector, where $\mathbf{d}_{\theta}$ denotes the learnable direction-reference network whose structure is again a combination of MHA and linear layers.
\begin{equation}
\begin{aligned}
\mathbf{v} = \dfrac{\mathbf{d}_{\theta}(\mathbf{f}_{0})}{\|\mathbf{d}_{\theta}(\mathbf{f}_{0})\|}. \quad 
\end{aligned}
\label{eq:unit}
\end{equation}
Next, the orthogonal component is computed as the perturbation direction and then sphere-projected onto the shell of radius $\varepsilon$, which can be expressed as follows:
\begin{equation}
\begin{aligned}
\boldsymbol{\delta}_{\parallel}= (\boldsymbol{\delta}^{\!\top}\mathbf{v})\mathbf{v},\qquad
\boldsymbol{\delta}_{\perp}= \varepsilon\dfrac{\boldsymbol{\delta}-\boldsymbol{\delta}_{\parallel}}{\|\boldsymbol{\delta}-\boldsymbol{\delta}_{\parallel}\|},
\end{aligned}
\end{equation}
where $\delta_{\parallel}$ and $\delta_{\perp}$ denote the parallel and orthogonal components, respectively.
Finally, we obtain the perturbed output of $\mathbf{f}_{0}$:
\begin{equation}
\begin{aligned}
\mathbf{f}_{\mathrm{adv}}= \mathbf{f}_{0}+\boldsymbol{\delta}_{\perp},\qquad
\end{aligned}
\end{equation}
By means of structured re-parameterization, direction and magnitude are disentangled, so that the constraint acts only in the insensitive orthogonal subspace. This keeps the key direction intact and stabilizes training.

\textbf{Reward function.} We define the reward function as follows:
\begin{equation}
\begin{aligned}
r(\mathbf{f}_{\mathrm{adv}},\mathbf{f}_{0},\mathbf{f}_{\mu})=
-\alpha\underbrace{\|\mathbf{f}_{\mathrm{adv}}-\mathbf{f}_{0}\|_{2}}_{\text{proximity}}
-\beta\underbrace{p_{\Phi}(\mathbf{f}_{\mathrm{adv}})}_{\text{PII probability}}\\
-\gamma\underbrace{\|\mathbf{f}_{\mathrm{adv}}-\mathbf{f}_{\mu}\|_{2}}_{\text{cluster prior}}.
\end{aligned}
\end{equation}
Among them, $p_{\Phi}(\cdot)$ represents the PII-classifier output, i.e., the probability that the adversarial feature contains personally identifiable information. Inspired by Wang et al. \cite{wang2024crafter}, $\mathbf{f}_\mu$ is centroid of the target cluster, used as a prior anchor, which is beneficial to hide PII. $\alpha$, $\beta$, $\gamma$ are the non-negative coefficients balancing proximity, privacy leakage, and cluster affinity in the reward.
Next, we define the baseline as follows.
\begin{equation}
\begin{aligned}
\mathbf{R}=r(\mathbf{f}_{\mathrm{adv}},\mathbf{f}_{0},\mathbf{f}_{\mu})\in\mathbb{R}^{B},
\end{aligned}
\end{equation}
\begin{equation}
\begin{aligned}
b_{t}=a\,b_{t-1}+c\,\bar{R},\quad \bar{R}=\mathbb{E}_{\text{batch}}[\mathbf{R}], \quad b_{0}=0,
\end{aligned}
\end{equation}
where a and c are the coefficients that regulate the baseline.
Then, compute the policy log-probability, which propagates the reward signal $\mathbf{R}$ backward along the sampled perturbation (single-step action) to update the policy-network parameters $\theta$.
\begin{equation}
\begin{aligned}
\log\pi_{\theta}(\mathbf{a}\mid\mathbf{s}) &= \sum_{i=1}^{D}\log\mathcal{N}(\delta_{i}\mid\mu_{i},\sigma_{i}^{2}).
\end{aligned}
\end{equation}
Therefore, the final objective is designed as follows and can be optimized via the above gradient.
\begin{equation}
\nabla_{\!\boldsymbol{\theta}}\mathcal{L}
= -\mathbb{E}_{\text{batch}}\!\Bigl[\bigl(\mathbf{R}-b_{t}\bigr)
\nabla_{\!\boldsymbol{\theta}}\log\pi_{\theta}(\mathbf{a}\mid\mathbf{s})\Bigr].
\end{equation}

\noindent
\begin{algorithm}[t]
\caption{TextCrafter.}
\label{alg:textCrafter}\small

   \KwIn{
      Encoder $E$, clean text $x$, policy network $\pi_{\theta}$,
      PII-classifier $p_{\phi}$, cluster-member count $N$,
      proximity coefficient $\alpha$, privacy coefficient $\beta$,
      cluster-affinity coefficient $\gamma$, projection-sphere radius $\varepsilon$
   }

   \KwOut{protected embedding $\mathbf{f}_{\mathrm{adv}}$}

   $\mathbf{f}_{0} \gets E(x)$;\quad $b_{0} \gets 0$;\quad
   $\mathbf{f}_{\mu} \gets \mathrm{mean}(N)$;

   \While{not meeting termination condition}{
      $\mathcal{N}(\boldsymbol{\mu},\sigma^{2}\mathbf{I}) \gets \mathrm{distribution}(\mathbf{f}_{0}+\mathbf{p})$;

      $\delta \gets$ sample from $\mathcal{N}(\boldsymbol{\mu},\sigma^{2}\mathbf{I})$;

      $\mathbf{v} \gets \mathrm{direction}(\mathbf{f}_{0}+\mathbf{p})$;

      $\log\pi_{\theta}(\mathbf{a}\mid\mathbf{s}) \gets \mathrm{prob}(\delta)$;

      $\delta_{\perp} \gets \mathrm{orthogonal}(\delta,\mathbf{v},\varepsilon)$;

      $\mathbf{f}_{\mathrm{adv}} \gets \mathbf{f}_{0} + \delta_{\perp}$;

      $\bar{R} \gets \mathrm{reward}(\mathbf{f}_{\mathrm{adv}},\mathbf{f}_{0},\mathbf{f}_{\mu},\alpha,\beta,\gamma,p_{\phi})$;

      $b_{t} \gets \mathrm{baseline}(\bar{R},b_{t-1})$;

      compute gradient $\nabla_{\!\boldsymbol{\theta}}\mathcal{L}$;

      update policy network and store best $\mathbf{f}_{\mathrm{adv}}^{*}$;
   }

   \Return{$\mathbf{f}_{\mathrm{adv}}^{*}$} \\
{\small\textbf{Notes.} \texttt{distribution} means construct the Gaussian action distribution $(\mu,\sigma^2)$;~
\texttt{direction} produces unit vector $\mathbf{v}$ of sensitive direction;~
\texttt{orthogonal} projects $\delta$ orthogonal to $\mathbf{v}$ and sphere-clips to radius $\varepsilon$;~
\texttt{reward} computes distance $+$ PII $+$ cluster reward;~
\texttt{baseline} is exponential moving average for variance reduction.}
\end{algorithm}

\begin{table*}[t]
\centering
\scriptsize
\caption{Privacy and Utility of Different Defense Methods across Datasets.}
\label{tab:defense_comparison}
\setlength{\tabcolsep}{0.95pt}
\renewcommand{\arraystretch}{1.05}
\begin{tabular}{l|cccccc|cccccc|cccccc|cccccc}
\toprule
\multirow{2}{*}{\textbf{Defense}} &
\multicolumn{6}{c|}{\textbf{Financial PhraseBank}} &
\multicolumn{6}{c|}{\textbf{ADE}} &
\multicolumn{6}{c|}{\textbf{SST-2}} &
\multicolumn{6}{c}{\textbf{AG News}}\\
\cmidrule{2-25}
 & BLEU$\downarrow$& T-F1$\downarrow$ & EM$\downarrow$ & R-L$\downarrow$& F1$\uparrow$ & Acc$\uparrow$
 & BLEU$\downarrow$ & T-F1$\downarrow$ & EM$\downarrow$ & R-L$\downarrow$ & F1$\uparrow$& Acc$\uparrow$ 
 & BLEU$\downarrow$ & T-F1$\downarrow$ & EM$\downarrow$ & R-L$\downarrow$ & F1$\uparrow$& Acc$\uparrow$
  & BLEU$\downarrow$ & T-F1$\downarrow$ & EM$\downarrow$ & R-L$\downarrow$ & F1$\uparrow$& Acc$\uparrow$\\
\midrule
No-Protection & 83.54 & 90.42 & 68.00 & 92.62 & 93.46 & 95.50 &
   91.08 & 95.22 & 84.00 & 95.99 & 96.93 & 98.00 &
   76.10 & 91.88 & 83.00 & 95.50 & 98.49 & 98.50  &
   59.07 & 82.44 & 32.00 & 77.99 & 85.47 & 86.00\\
 Shuffling&  1.25 &  7.47 &   0.00 &  8.10 & 11.22 & 12.00 & 
    0.86 &  5.22 &   0.00 &  5.25 & 44.40 & 51.00 & 
     1.20 &  7.85 &   0.00 &  7.46 & 22.20 & 31.00  & 
     1.69 &  6.45 &   0.00 &  7.91 & 32.00 & 44.00\\
 Gaussian ($\varepsilon=0.1$)    & 1.16 & 9.76 & 0.00 & 10.42 & 35.52 & 65.00 & 1.01 & 8.39 & 0.00 & 8.27 & 46.79 & 63.00 & 1.54 & 12.65 & 0.01 & 14.36 & 72.97 & 73.00 & 0.97 & 13.37 & 0.00 & 9.38 & 49.68 & 63.50\\
 LDP ($\varepsilon=4$)       & 3.70 & 4.56 & 0.00 & 5.07 & 27.28 & 63.00 & 9.10 & 4.46 & 0.00 & 4.42 & 44.44 & 62.00 & 7.20 & 4.09 & 0.00 & 3.55 & 48.00 & 52.50 & 5.75 & 20.54 & 0.00 & 6.69 & 27.50 & 59.50\\
 TextCrafter ($\varepsilon=2.5$)      & 2.70 & 9.14 & 0.00 & 13.67 & 68.0 & 76.00 & 2.10 & 9.75 & 0.01 & 11.50 & 50.87 & 75.00 & 1.24 & 16.49 & 0.00 & 11.68 & 68.00  & 79.00 & 1.88 & 16.63 & 0.00 & 7.43 & 65.10 & 74.50\\
 
\bottomrule
\end{tabular}
\end{table*}

\subsection{Theoretical Guarantees}

We now establish deterministic, sample-independent bounds that hold \emph{at inference time} for any input embedding $\mathbf{f}_{0}$.

Throughout this section we assume
\begin{itemize}
  \item $\|\mathbf{f}_{0}\|_{2}\le R_{0}$ with $R_{0}=\sup_{\mathbf{f}_{0}}\|\mathbf{f}_{0}\|_{2}$;
  \item the policy Gaussian $\pi_{\theta}(\cdot|\mathbf{s})=\mathcal{N}\bigl(\boldsymbol{\mu}_{\theta}(\mathbf{s}),\sigma^{2}\mathbf{I}\bigr)$ is \emph{fixed} (i.e.\ training has stopped);
  \item $\mathbf{v}=\mathbf{v}_{\theta}(\mathbf{s})\in\mathbb{R}^{D}$ is the deterministic unit reference direction;
  \item $\varepsilon>0$ is the user-chosen projection-sphere radius.
  \item $\alpha,\beta,\gamma>0$ are user-chosen constants.
\end{itemize}

\subsubsection{Perturbation Norm and Feature Drift}
\begin{lemma}[Orthogonal-component norm]
For any random vector $\boldsymbol{\delta}\sim\pi_{\theta}(\cdot|\mathbf{s})$ the perturbation
\begin{equation}
\boldsymbol{\delta}_{\perp}=\varepsilon\cdot\frac{\boldsymbol{\delta}-(\boldsymbol{\delta}^{\top}\mathbf{v})\mathbf{v}}{\|\boldsymbol{\delta}-(\boldsymbol{\delta}^{\top}\mathbf{v})\mathbf{v}\|_{2}}
\end{equation}
satisfies $\|\boldsymbol{\delta}_{\perp}\|_{2}=\varepsilon$ deterministically.
\end{lemma}

\begin{proof}
The numerator is orthogonal to $\mathbf{v}$, hence its $\ell_{2}$-norm is strictly positive.  Dividing by this norm yields a unit vector; scaling by the scalar $\varepsilon$ gives the claim.
\end{proof}

\begin{corollary}[Feature-space drift bound]
The protected embedding $\mathbf{f}_{\mathrm{adv}}=\mathbf{f}_{0}+\boldsymbol{\delta}_{\perp}$ satisfies
\begin{equation}
\|\mathbf{f}_{\mathrm{adv}}-\mathbf{f}_{0}\|_{2}=\varepsilon.
\end{equation}
Thus the maximum representation drift is \emph{independent of the sampled} $\boldsymbol{\delta}$ and is controlled \emph{solely} by the hyper-parameter $\varepsilon$. In other words, $\varepsilon$ serves as a deterministic privacy budget in the embedding space: the larger $\varepsilon$, the stronger the privacy protection, with the worst-case drift exactly capped at $\varepsilon$.
\end{corollary}

\subsubsection{Reward Lower Bound}
Recall the embedding-level reward
\begin{equation}
r(\mathbf{f}_{\mathrm{adv}})=-\alpha\|\mathbf{f}_{\mathrm{adv}}-\mathbf{f}_{0}\|_{2}-\beta p_{\Phi}(\mathbf{f}_{\mathrm{adv}})-\gamma\|\mathbf{f}_{\mathrm{adv}}-\mathbf{f}_{\mu}\|_{2},
\end{equation}
where $p_{\Phi}(\cdot)\in[0,1]$ is the PII-classifier probability and $\mathbf{f}_{\mu}$ is the target-cluster centroid.

\begin{lemma}[Reward lower bound]
For any $\mathbf{f}_{\mathrm{adv}}$ generated as above,
\begin{equation}
r(\mathbf{f}_{\mathrm{adv}})\ge-\alpha\varepsilon-\beta-\gamma(2R_{0}+\varepsilon).
\end{equation}
\end{lemma}

\begin{proof}
We bound each term separately:
\begin{enumerate}
  \item Proximity:\; $-\alpha\|\mathbf{f}_{\mathrm{adv}}-\mathbf{f}_{0}\|_{2}=-\alpha\varepsilon$.
  \item PII:\; $-\beta p_{\Phi}(\mathbf{f}_{\mathrm{adv}})\ge-\beta\cdot 1$.
  \item Cluster:\; by the triangle inequality
  \begin{align}
  \|\mathbf{f}_{\mathrm{adv}}-\mathbf{f}_{\mu}\|_{2}
  &\le\|\mathbf{f}_{\mathrm{adv}}-\mathbf{f}_{0}\|_{2}+\|\mathbf{f}_{0}-\mathbf{f}_{\mu}\|_{2}\\[2pt]
  &\le\varepsilon+\|\mathbf{f}_{0}\|_{2}+\|\mathbf{f}_{\mu}\|_{2}
  \le\varepsilon+2R_{0}.
  \end{align}
  Hence $-\gamma\|\mathbf{f}_{\mathrm{adv}}-\mathbf{f}_{\mu}\|_{2}\ge-\gamma(\varepsilon+2R_{0})$.
\end{enumerate}
Summing the three lower bounds yields the claim.
\end{proof}

\begin{corollary}[Monotonic privacy--utility trade-off]
The worst-case reward decreases \emph{linearly} with slope $-(\alpha+\gamma)$ as $\varepsilon$ increases.  Consequently, the privacy--utility balance is reduced to a \emph{single, monotonic and computable} knob.
\end{corollary}

\subsubsection{Summary of Inference-Time Guarantees}
\begin{theorem}
Let $\varepsilon,\alpha,\beta,\gamma>0$ be user-chosen constants.  For any input embedding $\mathbf{f}_{0}$ with $\|\mathbf{f}_{0}\|_{2}\le R_{0}$, TextCrafter outputs
\[
\mathbf{f}_{\mathrm{adv}}=\mathbf{f}_{0}+\boldsymbol{\delta}_{\perp},\quad\text{with}\quad\|\boldsymbol{\delta}_{\perp}\|_{2}=\varepsilon.
\]
Then deterministically
\begin{enumerate}
  \item $\|\mathbf{f}_{\mathrm{adv}}-\mathbf{f}_{0}\|_{2}=\varepsilon$ \emph{(utility drift)};  
  \item $\langle\mathbf{f}_{\mathrm{adv}}-\mathbf{f}_{0},\mathbf{v}\rangle=0$ \emph{(direction fidelity)};  
  \item $r(\mathbf{f}_{\mathrm{adv}})\ge-\alpha\varepsilon-\beta-\gamma(2R_{0}+\varepsilon)$ \emph{(privacy reward)}.
\end{enumerate}
All bounds are \emph{uniform} over the data distribution and require \emph{no re-training} at deployment.
\end{theorem}

These results certify that TextCrafter offers \emph{quantifiable, hyper-parameter-driven} privacy--utility guarantees at inference time. The schematic diagram of \textbf{TextCrafter} is shown in \reffig{fig:framework}.

\section{Results and Analysis}

\begin{figure}
\centering
	\captionsetup{
		font={scriptsize}, 
	}
	\begin{adjustbox}{valign=t}
		\includegraphics[width=1\linewidth]{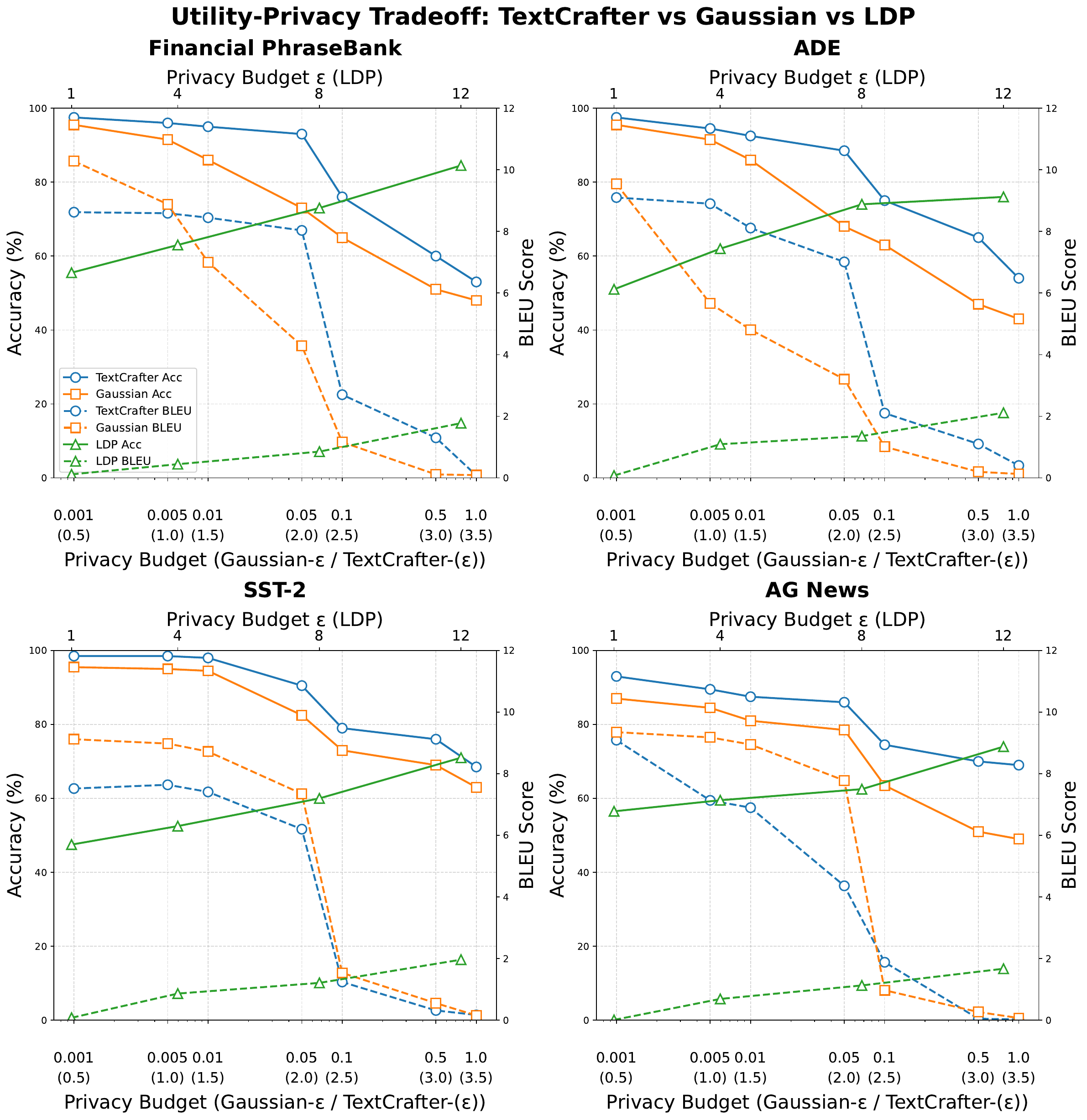}
	\end{adjustbox}
	\caption{Utility-Privacy Tradeoff Comparison across Different Datasets and Privacy Mechanisms. Bottom x-axis: Privacy budget $\epsilon$ for Gaussian/TextCrafter mechanisms; Top x-axis: Privacy budget $\epsilon$ for LDP mechanism. Left y-axis: Classification accuracy (\%) - higher values indicate better utility from defender's perspective; Right y-axis: BLEU score - lower values indicate stronger privacy (to a certain extent).}
	\label{fig:trade-off}
    \vspace{-4mm}
\end{figure}
After conducting comprehensive experiments on the Financial PhraseBank, ADE, SST-2, and AG News datasets. The key observation from \reftable{tab:defense_comparison} is that TextCrafter offers privacy comparable to Gaussian noisy injection—indicated to a certain extent by similar BLEU, Token-F1 and ROUGE-L values—yet delivers clearly superior downstream utility, improving accuracy by more than 10 percentage points. LDP ranks in the middle, while randomly shuffling embeddings yields the worst performance. \reffig{fig:trade-off} illustrates the privacy–utility trade-off of TextCrafter, Gaussian and LDP. Across all four datasets, TextCrafter consistently achieves a superior trade-off: it injects strong protective noise (manifested by the lower BLEU) while preserving high downstream utility (classification accuracy). This advantage stems from TextCrafter’s learnable perturbation mechanism, which constrains the noise to remain orthogonal to the sensitive sub-space, thereby averting large semantic shifts and safeguarding task-specific information.

\section{Conclusion}
This letter introduces \textbf{TextCrafter}, a learnable-perturbation method that trains a policy to trade off among the original embedding $\mathbf{f}_0$, cluster features $\mathbf{f}_u$, and PII signals, and then injects the perturbation orthogonal to $\mathbf{f}_0$ for better utility. Under strong privacy (\textsc{bleu}$<$3, \textsc{rouge-l}$<$15) it still maintains $\geq 70\%$ classification accuracy on four datasets, and outperforms Gaussian and LDP in most settings, including lower privacy budgets.

\balance

\bibliography{textCrafter.bib}

\begin{thebibliography}{10}
\providecommand{\url}[1]{#1}
\csname url@samestyle\endcsname
\providecommand{\newblock}{\relax}
\providecommand{\bibinfo}[2]{#2}
\providecommand{\BIBentrySTDinterwordspacing}{\spaceskip=0pt\relax}
\providecommand{\BIBentryALTinterwordstretchfactor}{4}
\providecommand{\BIBentryALTinterwordspacing}{\spaceskip=\fontdimen2\font plus
\BIBentryALTinterwordstretchfactor\fontdimen3\font minus \fontdimen4\font\relax}
\providecommand{\BIBforeignlanguage}[2]{{%
\expandafter\ifx\csname l@#1\endcsname\relax
\typeout{** WARNING: IEEEtran.bst: No hyphenation pattern has been}%
\typeout{** loaded for the language `#1'. Using the pattern for}%
\typeout{** the default language instead.}%
\else
\language=\csname l@#1\endcsname
\fi
#2}}
\providecommand{\BIBdecl}{\relax}
\BIBdecl

\bibitem{che2025llm}
H.~Che, H.~Jin, Z.~Gu, Y.~Lin, C.~Jin, and H.~Chen, ``Llm-driven medical report generation via communication-efficient heterogeneous federated learning,'' \emph{IEEE Transactions on Medical Imaging}, 2025.

\bibitem{appleSiri2025}
{Apple Inc.}, ``Siri personal assistant,'' \url{https://www.apple.com/siri/}, 2025, online; accessed 16-Sep-2025.

\bibitem{mahendran2015understanding}
A.~Mahendran and A.~Vedaldi, ``Understanding deep image representations by inverting them,'' in \emph{Proceedings of the IEEE conference on computer vision and pattern recognition}, 2015, pp. 5188--5196.

\bibitem{morris2023text}
J.~X. Morris, V.~Kuleshov, V.~Shmatikov, and A.~M. Rush, ``Text embeddings reveal (almost) as much as text,'' \emph{arXiv preprint arXiv:2310.06816}, 2023.

\bibitem{chen2024text}
Y.~Chen, H.~Lent, and J.~Bjerva, ``Text embedding inversion security for multilingual language models,'' \emph{arXiv preprint arXiv:2401.12192}, 2024.

\bibitem{dwork2014algorithmic}
C.~Dwork, A.~Roth \emph{et~al.}, ``The algorithmic foundations of differential privacy,'' \emph{Foundations and trends{\textregistered} in theoretical computer science}, vol.~9, no. 3--4, pp. 211--407, 2014.

\bibitem{du2023sanitizing}
M.~Du, X.~Yue, S.~S. Chow, and H.~Sun, ``Sanitizing sentence embeddings (and labels) for local differential privacy,'' in \emph{Proceedings of the ACM Web Conference 2023}, 2023, pp. 2349--2359.

\bibitem{jiang2023hexgen}
Y.~Jiang, R.~Yan, X.~Yao, Y.~Zhou, B.~Chen, and B.~Yuan, ``Hexgen: Generative inference of large language model over heterogeneous environment,'' \emph{arXiv preprint arXiv:2311.11514}, 2023.

\bibitem{presidio}
\BIBentryALTinterwordspacing
{Microsoft}, ``{Presidio: Context aware, pluggable and customizable PII anonymization service},'' GitHub repository, 2017, accessed: YYYY-MM-DD. [Online]. Available: \url{https://github.com/microsoft/presidio}
\BIBentrySTDinterwordspacing

\bibitem{vaswani2017attention}
A.~Vaswani, N.~Shazeer, N.~Parmar, J.~Uszkoreit, L.~Jones, A.~N. Gomez, {\L}.~Kaiser, and I.~Polosukhin, ``Attention is all you need,'' \emph{Advances in neural information processing systems}, vol.~30, 2017.

\bibitem{wang2024crafter}
S.~Wang, Z.~Ji, L.~Xiang, H.~Zhang, X.~Wang, C.~Zhou, and B.~Li, ``Crafter: Facial feature crafting against inversion-based identity theft on deep models,'' \emph{arXiv preprint arXiv:2401.07205}, 2024.

\end{thebibliography}
\bibliographystyle{IEEEtran}

\end{document}